\documentclass[
amsmath, %
floatfix, %
onecolumn, %
12pt, %
tightenlines, %
prb, %
aps, %
]{revtex4-2}

\usepackage[T1]{fontenc}
\usepackage[utf8]{inputenc}
\usepackage{graphicx}
\usepackage{latexsym}
\usepackage{amsthm}
\usepackage{color}
\usepackage{mathtools}
\usepackage{xcolor}
\usepackage{blkarray}

\usepackage[]{newtxtext}
\usepackage[nosymbolsc,smallerops,bigdelims]{newtxmath}
\DeclareMathAlphabet{\mathcal}{OMS}{cmsy}{m}{n}

\DeclareMathAlphabet{\mathcalb}{OMS}{cmsy}{b}{n}
\usepackage{bm}
\usepackage{hyperref}
\hypersetup{
	colorlinks,
	linkcolor={blue!90!black},
	citecolor={blue!90!black},
	urlcolor={blue!90!black}
}
\makeatletter
\renewcommand*{\eqref}[1]{%
	\hyperref[{#1}]{\textup{\tagform@{\ref*{#1}}}}%
}
\makeatother

\usepackage[cal=cm,calscaled=.95, %
            scr=boondoxo,scrscaled=1.05, %
            bb=boondox
            ]{mathalfa}

\newcommand{\mathbfbb}[1]{\bm{\mathbb{#1}}}

\DeclarePairedDelimiter\lr{\lparen}{\rparen}
\DeclarePairedDelimiter\Lr{\lbrack}{\rbrack}
\DeclarePairedDelimiter\LR{\lbrace}{\rbrace}

\DeclarePairedDelimiter\avg{\langle}{\rangle}

\DeclarePairedDelimiterX{\comm}[2]{\lbrack}{\rbrack}{#1, #2}
\DeclarePairedDelimiterX{\acomm}[2]{\lbrace}{\rbrace}{#1, #2}

\DeclarePairedDelimiter\ket{\lvert}{\rangle}
\DeclarePairedDelimiter\bra{\langle}{\rvert}

\DeclarePairedDelimiterX{\braket}[2]{\langle}{\rangle}{#1\delimsize\vert #2}
\DeclarePairedDelimiterX{\ketbra}[2]{\rvert}{\lvert}{#1 \delimsize\rangle\!\delimsize\langle #2}
\DeclarePairedDelimiterX{\matrixel}[3]{\langle}{\rangle}{#1 \delimsize\rvert #2 \delimsize\lvert #3}

\renewcommand{\tilde}[1]{\widetilde{#1}}

\newcommand{\ii}{\mathrm{i}}
\newcommand{\e}{\mathrm{e}}

\newcommand{\lkaka}{\lambda_{k\alpha k'\alpha'}}
\newcommand{\nka}{\nu_{k\alpha}}

\newcommand{\Trpq}{\mathrm{Tr}_{pq}}
\newcommand{\Rkk}{\mathcal{R}_{kk'}}
\newcommand{\Ukk}{\mathcal{U}_{kk'}}

\newcommand{\JOBM}{\frac{J_0\beta}{M}}
\newcommand{\JBM}{\frac{J\beta}{M}}

\newcommand{\Si}{\sum_i}

\newcommand{\aaa}{\alpha\alpha'}

\newcommand{\mm}[2]{\mu^{(#1)}_{#2}}
\newcommand{\nminus}[1]{\mbox{\footnotesize $(n{-}#1)$}}

\newcommand{\vecel}[1]{\mathbb{#1}}
\newcommand{\intid}{\int\limits_{-\infty}^{\infty}\!\!\mathrm{d}}

\newcommand{\vrho}{\rho_{\alpha k}         }
\newcommand{\veps}{\epsilon_{\alpha kk'}   }
\newcommand{\veta}{\eta_{\alpha\alpha' kk'}}
\newcommand{\vzet}{\zeta_{\alpha kk'}      }
\newcommand{\vxxi}{\xi_{\alpha\alpha' kk'} }

\begin{document}
\title{Stability of the replica-symmetric solution in the off-diagonally-disordered Bose-Hubbard model}
\author{Anna M. Piekarska}
\author{Tadeusz K. Kope\'{c}}
\affiliation{Institute of Low Temperature and Structure Research, Polish Academy of Sciences, 
Ok\'{o}lna 2, 50-422 Wroc\l{}aw, Poland}
\begin{abstract}
We study a disordered system of interacting bosons described by the Bose-Hubbard Hamiltonian with random tunneling amplitudes.
We derive the condition for the stability of the replica-symmetric solution for this model.
Following the scheme of de Almeida and Thouless, we determine if the solution corresponds to the minimum of free energy by building the respective Hessian matrix and checking its positive semidefiniteness.
Thus, we find the eigenvalues by postulating the set of eigenvectors based on their expected symmetry, and require the eigenvalues to be non-negative.
We evaluate the spectrum numerically and identify matrix blocks that give rise to eigenvalues that are always non-negative.
Thus, we find a subset of eigenvalues coming from decoupled subspaces that is sufficient to be checked as the stability criterion.
We also determine the stability of the phases present in the system, finding that the disordered phase is stable, the glass phase is unstable,
while the superfluid phase has both stable and unstable parts.
\end{abstract}
\maketitle
\tableofcontents
%
\section{Introduction}
Investigation of strongly interacting disordered bosonic systems, enabled by recent progress in quantum simulation,
allows for a revival of rich physics found in the field of spin glasses, but this time with an added ingredient of strong interactions.

Spin glasses are frustrated magnetic systems that have attracted much interest in the past
as they feature unique properties that were found to be hard to be fully explained~\cite{Binder1986_RoMP58}.
Recognizing the spin-glass phase requires averaging over the quenched randomness,
which can be quantified by the Edwards-Anderson order parameter~\cite{Edwards1975_JoPFMP5}.
A mean-field theory of spin glasses was proposed in a seminal work by Sherrington and Kirkpatrick~\cite{Sherrington1975_PRL35}.
Their solution of the proposed model employed the so-called replica trick in order to deal with the quenched average.

There was, however, a problem with this solution: the entropy of the glass phase was negative.
The origin of the problem was the assumption that the solution should be symmetric in the replica space.
The stability of the replica-symmetric solution has been analyzed in detail by de Almeida and Thouless~\cite{Almeida1978_JoPAMaG11}.
They found that it is stable in the entire disordered phase but unstable in the whole glass phase.
Subsequently, the stability has been studied in the quantum case~\cite{Ray1989_PRB39,Thirumalai1989_JoPAMaG22,Buettner1990_PRB41,Goldschmidt1990_PRL64}.
The outcomes were not consistent: the analysis using Monte Carlo simulations~\cite{Ray1989_PRB39}
and the one using the static approximation~\cite{Thirumalai1989_JoPAMaG22} indicated that some part of the spin-glass phase is stable.
In contrast, the studies which went beyond the static approximation~\cite{Buettner1990_PRB41,Goldschmidt1990_PRL64}
concluded that the whole glass phase is unstable, as it was in the classical case.
In systems where more phases were present, and the usual Edwards-Anderson order was not enough to distinguish the glassy phase,
stability of the solution has been used as an indication of glassiness~\cite{Kopec1990_JoPCM2,Ma1991_JoPCM3,Yu2012_PRB85}.
%

When a disordered system is also strongly interacting, these two features compete,
leading to the emergence of interesting physical phenomena~\cite{Giamarchi1988_PRB37}.
A new playground for investigating the properties of such systems came with the development of quantum simulators~\cite{Bloch2008_RoMP80}.
In particular, in bosonic systems, one can study the influence of disorder on the superfluid state~\cite{Gimperlein2005_PRL95},
which is an important step in understanding phenomena like high-$T_\mathrm{c}$ superconductivity~\cite{Bednorz1986_ZfPBCM64}
or supersolidity in ${}^{4}$He~\cite{Hunt2009_S324}.
Most of the studies of disordered bosonic systems focus on diagonal kind of disorder~\cite{Fisher1989_PRB40,Gurarie2009_PRB80},
i.e., where randomness is in the chemical potential term.
In such a case, a Bose glass emerges.
Although diagonal disorder does not introduce frustration, some quantities from the spin-glass world
like Edwards-Anderson (EA) order parameter~\cite{Thomson2014_EL108} (however, based on particle number density fluctuations)
or replica overlap~\cite{Morrison2008_NJoP10} can be found in the Bose glass systems.

In order to fully unravel the spin-glass features like slow relaxation, one needs to introduce frustration to the system.
It can be done by choosing the less explored case of an off-diagonal kind of disorder.
It has been found that frustrated Coulomb interactions lead to the emergence of a superglass~\cite{Yu2012_PRB85},
again with the EA order parameter defined based on local boson density.
In the case of frustrated hopping~\cite{Pazmandi1995_PRL74,Pazmandi1996,Piekarska2018_PRL120,Piekarska2022},
the EA order parameter of the resulting glassy phase is based on the complex $U$(1) phase of bosons.
Here, we focus on the latter.

In this paper, we study the stability of the replica-symmetric solution
in a system of interacting bosons with random tunnelings of non-zero mean,
described by the Bose-Hubbard model.
We derive a stability criterion following the method of de Almeida and Thouless~\cite{Almeida1978_JoPAMaG11}
and incorporating a way to deal with the Trotter space, which was used in a quantum spin-glass system~\cite{Buettner1990_PRB41}.
We find that the whole disordered phase is stable, while the entire glassy phase is unstable, in agreement with the aforementioned derivation.
Moreover, we find that the superfluid phase contains both stable and unstable areas.

The text is organized as follows.
First, in Section~\ref{sec:prelim}, we introduce the main building blocks of the further derivation.
Then, in Section~\ref{sec:trotter}, we introduce the Hessian matrix for free energy in our problem
and simplify its Trotter dimensions as much as possible, leaving a problem similar to that of the SK model.
Next, in Section~\ref{sec:matrixel}, we formulate the simplified problem in a manner suitable for the eigenvector postulation,
which we then perform in Section~\ref{sec:eigenvectors}.
In Section~\ref{sec:numerical}, we apply the conditions derived in previous sections to exemplary cross-sections of the parameter space
and use it to analyze the stability of various phases.
For readers' convenience, we give the statement of the simplified stability criterion in a separate Section~\ref{sec:criterion}.
Finally, in Section~\ref{sec:summary}, we summarize the findings.

\section{Preliminaries}\label{sec:prelim}
\subsection{Model}\label{sec:model}
The Bose-Hubbard Hamiltonian for the system of $N$ interacting bosons reads
\begin{equation}
	H = -\sum_{i<j}J_{ij}\lr*{a_i^\dagger a_j+a_j^\dagger a_i}
	    + \frac{U}{2}\Si \hat{n}_i\lr*{\hat{n}_i-1} - \mu \Si \hat{n}_i,
\end{equation}
where $a_i$ ($a_i^\dagger$) are the annihilation (creation) operators for the site $i$
and $\hat{n}_i=a_i^\dagger a_i$ are the particle number operators,
while $\mu$ and $U$ denote the chemical potential and on-site interaction strength, respectively.
$J_{ij}$ are the tunneling amplitudes (hopping integrals) between sites $i$ and $j$.
Here, we are interested in off-diagonally disordered model, in which $J_{ij}$ are independent random variables
given by Gaussian distribution with mean $J_0/N$ and variance $J^2/N$,
following Ref.~\cite{Sherrington1975_PRL35}.

Below, we briefly summarize the derivation for reader's convenience.
To obtain the effective free energy, we follow a scheme~\cite{Piekarska2022}
analogous to the one used in the quantum version~\cite{Bray1980_JoPCSSP13} of the Sherrington-Kirkpatrick model~\cite{Sherrington1975_PRL35}.
We start with the replica trick
\begin{equation}
	F = \frac{1}{\beta}\Lr*{\mathrm{ln} Z}_{J} = \lim_{n\to 0} \frac{1}{n\beta}\lr[\Big]{\Lr*{Z^{n}}_{J}-1},
\end{equation}
where $F$ is the free energy, $Z = \mathrm{Tr} \exp(-\beta H)$ is the partition function, $\beta = k_{\mathrm{B}}T$ and $\Lr*{\cdots}_{J}$ denotes the average over the disorder.
Lack of the logarithm in the expression later allows us to perform the integration over the disorder distribution.
Next, we transform the Hamiltonian to the quasi-momentum and quasi-positions basis,
\begin{equation}
	\hat{P} = \frac{i}{\sqrt{2}}\lr*{a^\dagger-a}, \qquad\hat{Q} = \frac{1}{\sqrt{2}}\lr*{a^\dagger+a},
\end{equation}
and handle the non-commuting terms of the Hamiltonian via the Trotter-Suzuki expansion~\cite{Suzuki1976_CiMP51},
i.e., we split the Hamiltonian into parts:
\begin{equation}
	H_P = -\sum_{i < j} J_{ij}\hat{P}_{i}\hat{P}_{j}, \quad
	H_Q = -\sum_{i < j} J_{ij}\hat{Q}_{i}\hat{Q}_{j}, \quad
	H_n = \frac{U}{2}\sum_{i}\hat{n}_{i}^2-\lr*{\mu+\frac{U}{2}}\sum_{i}\hat{n}_{i},
\end{equation}
and use the formula
\begin{equation}
	\exp\lr*{-\beta \lr*{H_{P}+H_{Q}+H_{n}}} = \lim_{M\to\infty} \Lr*{
	\exp\lr*{-\frac{\beta H_{P}}{M}} \exp\lr*{-\frac{\beta H_{n}}{2M}} \exp\lr*{-\frac{\beta H_{Q}}{M}} \exp\lr*{-\frac{\beta H_{n}}{2M}} }^{M}.
\end{equation}
Then, we insert the sums of projections $\sum_{x} \ket{x}\bra{x} = \mathbb{1}$ over a complete basis of eigenvectors of either $\hat{P}$ or $\hat{Q}$
between each pair of consecutive exponents, and get a classical expression
\begin{equation}
	\exp\lr*{-\beta H} = \lim_{M\to\infty} \Trpq\mathcal{M}\prod_{k=1}^{M}\prod_{\alpha=1}^{n}
	\exp\lr*{ \frac{\beta}{M}\sum_{i < j} J_{ij} \lr{ p_{ik}^{(\alpha)}p_{jk}^{(\alpha)} + q_{ik}^{(\alpha)}q_{jk}^{(\alpha)}}},
\end{equation}
with the quantum nature remaining only in the term
\begin{equation}
	\mathcal{M} = \prod_{i,k,\alpha} \matrixel*{p_{ik}^{(\alpha)}}{\e^{-\frac{\beta H_{n}}{2M}}}{q_{ik}^{(\alpha)}} \matrixel*{q_{ik}^{(\alpha)}}{\e^{-\frac{\beta H_{n}}{2M}}}{p_{i,k+1}^{(\alpha)}},
\end{equation}
which we will treat as an ad-hoc-defined function throughout the rest of the derivation.
In the above equations, we have introduced the eigenvalues and the corresponding eigenvectors of $\hat{P}$ and $\hat{Q}$,
i.e., $\hat{P}\ket{p_{ik}^{(\alpha)}} = p_{ik}^{(\alpha)}\ket{p_{ik}^{(\alpha)}}$,
while the trace $\Trpq$ runs over all of the combinations of these eigenvalues at all valid indices.
At this point, we carry out the integration over the disorder, which results in replica coupling.
To be able to take the thermodynamic limit, we apply the Hubbard-Stratonovich transformation to the terms that couple sites, for example,
\begin{equation}
	\exp\Lr*{
		\frac{1}{4N}\lr*{
			\JBM\Si p_{ik}^{(\alpha)}p_{ik'}^{(\alpha')}
		}^2
	}
	= \sqrt{\frac{N}{\pi}} \intid\lkaka \exp\Lr*{
		-N\lkaka^2+\lkaka\lr*{
			\JBM\Si p_{ik}^{(\alpha)}p_{ik'}^{(\alpha')}
		}
	}.
\end{equation}
In this step, a few sets of auxiliary fields are introduced.
Apart from the $\lkaka$ presented above, which are coupled to $p_{k\alpha}p_{k'\alpha'}$ or $q_{k\alpha}q_{k'\alpha'}$,
there are also $\lkaka^{(PQ)}$ coupled to $p_{k\alpha}q_{k'\alpha'}$ and $\nka$ coupled to $p_{k\alpha}$.
In the saddle point method, we obtain the effective free energy
\begin{equation}\label{eq:freeene}
	\mathcal{F} = 
		  \sum_{k\alpha k'\alpha'}\Lr*{2\lkaka^{2}+\frac{1}{2}\lr*{\lkaka^{PQ}}^{2}}
		+ \sum_{k\alpha}\nka^2
		- \ln \mathrm{Tr} \e^{-\beta \mathcal{H}},
\end{equation}
where
\begin{multline}
	-\beta \mathcal{H} = \JBM\sum_{k\alpha k'\alpha'}\Lr*{
		\lkaka\lr*{p_{k\alpha}p_{k'\alpha} + q_{k\alpha}q_{k'\alpha}}
		+ \lkaka^{(PQ)}p_{k\alpha}q_{k'\alpha}
	}\\
	+ \sqrt{\JOBM} \sum_{k\alpha} \nka \lr*{ p_{k\alpha} + q_{k\alpha}} + \ln \mathcal{M}.
\end{multline}
The variables $\lkaka$, $\lkaka^{(PQ)}$ and $\nka$ are defined self-consistently as
\begin{subequations}
\begin{eqnarray}
	\lkaka &=& \frac{J\beta}{2M}\avg{p_{k\alpha}p_{k'\alpha'}},\\
	\lkaka^{PQ} &=& \JBM\avg{p_{k\alpha}q_{k'\alpha'}},\\
	\nka &=& \sqrt{\JOBM}\avg{p_{k\alpha}}.
\end{eqnarray}
\end{subequations}

\subsection{Stability criterion}
In the described derivation, one would like to assume that the solution is symmetric in the replica space.
This simplification is known to be correct in the disordered phase,
but inaccurate in the glass phase \cite{Almeida1978_JoPAMaG11}.
The problem is that in the GL phase,
the saddle point solution is not the minimum of the free energy anymore.
While the situation is clear in the DI and GL phases,
the stability of the solution in the SF phase, that emerges when $J_{0}\neq 0$, needs investigation.
In the following, we find that SF has both stable and unstable parts.
The analysis of the properties of the latter allowed to classify it as a superglass phase~\cite{Piekarska2022},
similarly to a case of a different disorder inclusion~\cite{Yu2012_PRB85}.

We look for the instability following the scheme introduced by de Almeida and Thouless~\cite{Almeida1978_JoPAMaG11}.
To this end, we expand the free energy~\eqref{eq:freeene} around the replica-symmetric solution (characterized by $\varDelta$, $q$, $u$, $\Rkk$, $\Ukk$):
\begin{subequations}
\begin{align}
	\label{eq:def-vars-a}
	\nu_{k\alpha}                      &= \sqrt{\JOBM}    \lr*{\varDelta + \vrho },\\
	\lambda_{k\alpha k'\alpha}         &= \frac{1}{2}\JBM \lr*{\Rkk      + \veps },\\
	\lambda_{k\alpha k'\alpha'}        &= \frac{1}{2}\JBM \lr*{q         + \veta },\\
	\lambda^{(PQ)}_{k\alpha k'\alpha}  &=            \JBM \lr*{\Ukk      + \vzet },\\
	\label{eq:def-vars-e}
	\lambda^{(PQ)}_{k\alpha k'\alpha'} &=            \JBM \lr*{u         + \vxxi },
\end{align}
\end{subequations}
where $\rho, \epsilon, \eta, \zeta$ and $\xi$ are the replica-dependent deviations,
and build a Hessian matrix $G$ of second derivatives of $\mathcal{F}$
with respect to all these deviations.
This matrix needs to be positive semi-definite
for the replica-symmetric solution to give the free energy minimum.
Otherwise, no replica-symmetric solution exists, and a theory accounting for breaking this symmetry is needed.
We will analyze this matrix to find its lowest eigenvalue,
the sign of which will indicate whether the matrix is positive semi-definite or not.

At first, we will view the matrix $G$ as a matrix indexed by sets of Trotter indices,
with its elements being matrices indexed by replica indices.
We will use the translational invariance in the Trotter space
and Fourier transform $G$ to decouple it into $M$ independent blocks that have a common structure.
We will also notice that this transformation
introduces an additional decoupling within each block.
Then, we will consider each of the blocks separately.
We will reorder the labeling of the matrix elements
and, this time, view $G$ as indexed by replica indices,
with elements being matrices (or vectors/scalars) in the reciprocal Trotter space.
At this point, the structure of each independent block
will consist only of parts that have a counterpart in the AT matrix.
Thus, we will find its eigenvalues
by checking three kinds of eigenvectors of analogous structure as in the AT derivation:
symmetric in the replica space,
symmetric under interchange of all but one replica index,
symmetric under interchange of all but two replica indices.

\subsection{Second-order expansion}
The matrix $G$ consists of second derivatives of $\mathcal{F}$ with respect to variables $\rho, \epsilon, \eta, \zeta$ and $\xi$.
Let us derive the expressions for these derivatives.
Upon expanding according to Eqs.~\eqref{eq:def-vars-a}-\eqref{eq:def-vars-e}, the free energy has the form
\begin{equation}
	\mathcal{F} = \mathcal{F}^{(0)} + \mathcal{F}^{(1)} + \mathcal{F}^{(2)} - \ln \mathrm{Tr} \e^{-\beta \mathcal{H}}
\end{equation}
with
\begin{equation}
	-\beta \mathcal{H} = -\beta \mathcal{H}^{(0)} -\beta \mathcal{H}^{(1)},
\end{equation}
where we group terms of various orders in deviations from the replica-symmetric solution: 
\begin{subequations}
\begin{align}
	\mathcal{F}^{(0)} &= \frac{1}{2}\lr*{\JBM}^{2}\sum_{kk'}\Lr*{
		 n\lr*{\Rkk^{2}+\Ukk^{2}}
		+n(n-1)\lr*{q^{2}+u^{2}}
	} + \JOBM\sum_{k\alpha}\nka^{2},\\
	\mathcal{F}^{(1)} &= \lr*{\JBM}^{2}\sum_{kk'}\Lr*{
		  \sum_{\alpha}\lr*{\Rkk\veps+\Ukk\vzet}
		+2\sum_{\alpha<\alpha'}\lr*{q\veta+u\vxxi}
	} + 2\JOBM\sum_{k\alpha}\nka \vrho,\\
	\mathcal{F}^{(2)} &= \frac{1}{2}\lr*{\JBM}^{2}\sum_{kk'}\Lr*{
		 \sum_{\alpha}\lr*{\veps^{2}+\vzet^{2}}
		 +2\sum_{\alpha<\alpha'}\lr*{\veta^{2}+\vxxi ^{2}}
	} + \JOBM\sum_{k\alpha}\vrho^{2},
\end{align}
\end{subequations}
\begin{subequations}
\begin{multline}
	-\beta \mathcal{H}^{(0)} = \lr*{\JBM}^{2}\sum_{kk'}\sum_{\alpha<\alpha'}\Lr*{
		  q\lr*{p_{k\alpha}p_{k'\alpha'} + q_{k\alpha}q_{k'\alpha'}}
		+ u\lr*{p_{k\alpha}q_{k'\alpha'} + q_{k\alpha}p_{k'\alpha'}}
	}\\
	+\frac{1}{2}\lr*{\JBM}^{2}\sum_{kk'}\sum_{\alpha}\Lr*{
		  \Rkk\lr*{p_{k\alpha}p_{k'\alpha} + q_{k\alpha}q_{k'\alpha}}
		+2\Ukk p_{k\alpha}q_{k'\alpha}
	}\\
	+\JOBM \sum_{k\alpha} \varDelta \lr*{ p_{k\alpha} + q_{k\alpha}} + \ln \mathcal{M},
\end{multline}
\begin{multline}
	-\beta \mathcal{H}^{(1)} = \lr*{\JBM}^{2}\sum_{kk'}\sum_{\alpha<\alpha'}\Lr*{
		  \veta\lr*{p_{k\alpha}p_{k'\alpha'} + q_{k\alpha}q_{k'\alpha'}}
		+ \vxxi\lr*{p_{k\alpha}q_{k'\alpha'} + q_{k\alpha}p_{k'\alpha'}}
	}\\
	+\frac{1}{2}\lr*{\JBM}^{2}\sum_{kk'}\sum_{\alpha}\Lr*{
		  \veps\lr*{p_{k\alpha}p_{k'\alpha} + q_{k\alpha}q_{k'\alpha}}
		+2\vzet p_{k\alpha}q_{k'\alpha}
	}\\
	+\JOBM \sum_{k\alpha} \vrho \lr*{ p_{k\alpha} + q_{k\alpha}} + \ln \mathcal{M},
\end{multline}
\end{subequations}
Second derivatives of $\mathcal{F}$ come entirely from the terms $\mathcal{F}^{(2)}$ and $-\beta \mathcal{H}^{(1)}$
and have a general form
\begin{equation}\label{eq:deri}
	\frac{\partial^{2} \mathcal{F}}{\partial u\partial v} = \mathcal{D}_{uv}
	+ \avg*{\frac{\partial(-\beta \mathcal{H})}{\partial u}}\avg*{\frac{\partial(-\beta \mathcal{H})}{\partial v}}
	- \avg*{\frac{\partial(-\beta \mathcal{H})}{\partial u}\frac{\partial(-\beta \mathcal{H})}{\partial v}},
\end{equation}
where $\mathcal{D}_{uv}$ is the diagonal element given by
\begin{equation}
	\mathcal{D}_{uv} = \begin{cases}
		0               & \mbox{for } u\neq v\\
		2\JOBM          & \mbox{for } u=v=\vrho\\
		\lr*{\JBM}^{2}  & \mbox{for } u=v=\veps \mbox{ or } u=v=\vzet\\
		2\lr*{\JBM}^{2} & \mbox{for } u=v=\eta_{\aaa kk'} \mbox{ or } u=v=\xi_{\aaa kk'}
	\end{cases},
\end{equation}
and the derivatives of the Hamiltonian are
\begin{subequations}
\begin{align}
	\label{eq:ham-derivs-a}
	\frac{\partial(-\beta \mathcal{H})}{\partial \vrho} &=                \JOBM    \lr[\Big]{p_{\alpha k}+q_{\alpha k}},                      \\
	\frac{\partial(-\beta \mathcal{H})}{\partial \veps} &= \frac{1}{2}\lr*{\JBM}^2 \lr[\Big]{p_{\alpha k}p_{\alpha k'}+q_{\alpha k}q_{\alpha k'}},\\
	\frac{\partial(-\beta \mathcal{H})}{\partial \vzet} &= \frac{1}{2}\lr*{\JBM}^2 \lr[\Big]{p_{\alpha k}q_{\alpha k'}+q_{\alpha k}p_{\alpha k'}},\\
	\frac{\partial(-\beta \mathcal{H})}{\partial \veta} &=            \lr*{\JBM}^2 \lr[\Big]{p_{\alpha k}p_{\alpha'k'}+q_{\alpha k}q_{\alpha'k'}},\\
	\label{eq:ham-derivs-e}
	\frac{\partial(-\beta \mathcal{H})}{\partial \vxxi} &=            \lr*{\JBM}^2 \lr[\Big]{p_{\alpha k}q_{\alpha'k'}+q_{\alpha k}p_{\alpha'k'}}.
\end{align}
\end{subequations}

\subsection{Useful identities}
Throughout the derivation, we utilize typical rules
of evaluating averages containing replicas and Trotter indices, i.e.:
\begin{enumerate}
	\item Replicas are uncorrelated,
		\begin{equation}\label{eq:rule-repl-uncorrel}
			\avg*{\mathscr{f}(\alpha)\mathscr{g}(\alpha')} = \avg*{\mathscr{f}(\alpha)}\avg*{\mathscr{g}(\alpha')}.
		\end{equation}
	\item Averages do not depend on the replica index,
		\begin{equation}\label{eq:rule-repl-indep}
			\avg*{\mathscr{f}(\alpha)} = \avg*{\mathscr{f}(\alpha')} \equiv \avg*{\mathscr{f}}.
		\end{equation}
	\item Averages do not depend on common translations in the Trotter space
	(but relative differences are important),
		\begin{equation}\label{eq:rule-trot-transl}
			\avg*{\mathscr{f}(k)\mathscr{g}(k+\Delta k)} = \avg*{\mathscr{f}(k+k_{0})\mathscr{g}(k+k_{0}+\Delta k)}.
		\end{equation}
	\item Separate averages do not depend on relative differences in the Trotter space,
		\begin{equation}\label{eq:rule-trot-relat}
			\avg*{\mathscr{f}(k)}\avg*{\mathscr{g}(k+\Delta k)} = \avg*{\mathscr{f}(k)}\avg*{\mathscr{g}(l)}.
		\end{equation}
\end{enumerate}
In the above, $\mathscr{f}(\alpha)$ denotes any expression
containing one or more variables $p$ or $q$
with replica index $\alpha$ and no other replica indices,
e.g. $\mathscr{f}(\alpha) = p_{\alpha k}q_{\alpha k'}$.
Analogously, $\mathscr{f}(k)$ denotes any expression with Trotter index $k$.

\section{Simplifying Trotter dimensions}\label{sec:trotter}
We start with reducing the number of Trotter dimensions in the matrix.
We follow Ref.~\cite{Buettner1990_PRB41},
where a quantum Ising spin glass was studied in a similar manner,
and perform Fourier transformation in the Trotter space.
It acts on the matrix, which has the structure
\begin{equation}\label{eq:struct-ka}
	\Lr*{\Lr*{G_{uv}}_{\lr{k,k+\Delta k}, \lr{k+l,k+l+\Delta l}}}_{\LR{\alpha},\LR{\beta}} =
	\frac{\partial^{2} \mathcal{F}}{\partial u_{\LR{\alpha},k,k+\Delta k} \partial v_{\LR{\beta},k+l,k+l+\Delta l}},
\end{equation}
where $u,v \in \LR{\rho, \epsilon, \zeta, \eta, \xi}$ denote blocks within the matrix,
corresponding to differentiation variables.
They are further subdivided into subblocks indexed by pairs of Trotter indices
(for $u=\rho$ or $v=\rho$ the second Trotter index is absent).
Finally, elements of these subblocks are indexed by sets $\LR{\alpha}, \LR{\beta}$ of one or two replica indices,
where the number of indices depends on the block, as can be seen from Eqs.~\eqref{eq:def-vars-a}-\eqref{eq:def-vars-e}.
We Fourier transform $G$,
\begin{multline}
	\Lr*{\Lr*{G_{uv}}_{(x,y), (x',y')}}_{\LR{\alpha},\LR{\beta}} =\\[5pt]
	\sum_{k,\Delta k,l,\Delta l} \e^{\ii kx}\e^{\ii(k+\Delta k)y}\e^{\ii(k+l)x'}\e^{\ii(k+l+\Delta l)y'}
	\Lr*{\Lr*{G_{uv}}_{\lr{k,k+\Delta k}, \lr{k+l,k+l+\Delta l}}}_{\LR{\alpha},\LR{\beta}},
\end{multline}
introducing the reciprocal space indices $x$, $x'$, $y$, and $y'$
dual to respective linear combinations of Trotter indices.
Since all the terms in $\partial^{2}\mathcal{F}/\partial u\partial v$
are invariant upon translations in the Trotter space,
we expect this matrix to be decoupled into independent blocks.
To check this, we analyze which elements vanish
by checking which Trotter indices they depend on
or, more importantly, on which they do not depend.

In this section, we are only interested in the dependence on Trotter indices,
so we consider replica-indexed matrices $\Lr*{G_{uv}}_{\lr{k,k+\Delta k}, \lr{k+l,k+l+\Delta l}}$ as a whole.
Combining rules~\eqref{eq:rule-repl-uncorrel} and~\eqref{eq:rule-trot-relat}, one can see
that when two expressions are indexed with different replicas,
the average of their product does not depend on the relative difference of their Trotter indices.
Thus, among the elements of a matrix $\Lr*{G_{uv}}_{\lr{k,k+\Delta k}, \lr{k+l,k+l+\Delta l}}$,
those with the most repeated replica indices (so diagonal elements)
will depend on at least as many Trotter indices as other elements of the same matrix.
Since we are checking whether the whole matrix $\Lr*{G_{uv}}_{(x,y), (x',y')}$ vanishes,
it is sufficient to analyze only the replica-diagonal elements.
Throughout this section, we will drop replica indices and treat the expressions
$\Lr*{G_{uv}}_{\lr{k,k+\Delta k}, \lr{k+l,k+l+\Delta l}}$ and $\Lr*{G_{uv}}_{(x,y), (x',y')}$
as selected diagonal elements.

Let us analyze contributions to $\Lr*{G_{uv}}_{(x,y), (x',y')}$
coming from various terms of $\partial^{2}\mathcal{F}/\partial u\partial v$.
According to Eq.~\eqref{eq:deri},
$\Lr*{\Lr*{G_{uv}}_{\lr{k,k+\Delta k}, \lr{k+l,k+l+\Delta l}}}_{\LR{\alpha},\LR{\beta}}$ consists of three terms which read
\begin{subequations}
\begin{align}
	\label{eq:def-G1}
	\Lr*{\Lr*{G_{uv}}_{\lr{k,k+\Delta k}, \lr{k+l,k+l+\Delta l}}}_{\LR{\alpha},\LR{\beta}}^{(1)} &=
		\mathcal{D}_{u_{\LR{\alpha},k,k+\Delta k}v_{\LR{\beta},k+l,k+l+\Delta l}},\\
	\Lr*{\Lr*{G_{uv}}_{\lr{k,k+\Delta k}, \lr{k+l,k+l+\Delta l}}}_{\LR{\alpha},\LR{\beta}}^{(2)} &=
		\avg*{\frac{\partial(-\beta \mathcal{H})}{\partial u_{\LR{\alpha},k,k+\Delta k}}}\avg*{\frac{\partial(-\beta \mathcal{H})}{\partial v_{\LR{\beta},k+l,k+l+\Delta l}}},\\
	\label{eq:def-G3}
	\Lr*{\Lr*{G_{uv}}_{\lr{k,k+\Delta k}, \lr{k+l,k+l+\Delta l}}}_{\LR{\alpha},\LR{\beta}}^{(3)} &=
		\avg*{\frac{\partial(-\beta \mathcal{H})}{\partial u_{\LR{\alpha},k,k+\Delta k}}\frac{\partial(-\beta \mathcal{H})}{\partial v_{\LR{\beta},k+l,k+l+\Delta l}}}.
\end{align}
\end{subequations}
Thus, $\Lr*{G_{uv}}_{(x,y), (x',y')}$ also splits into three terms,
being Fourier transforms of Eqs.~\eqref{eq:def-G1}-\eqref{eq:def-G3}.

\subsection{Diagonal contributions}
The first term is non-zero only when $u_{\LR{\alpha},k,k+\Delta k}=v_{\LR{\beta},k+l,k+l+\Delta l}$.
In particular, $u=v$ must hold.
We consider each of the blocks separately.
For example, in the $\epsilon\epsilon$ block, we have
\begin{multline}
	\Lr*{G_{\epsilon\epsilon}}_{(x,y), (x',y')}^{(1)} = \sum_{k,\Delta k,l,\Delta l} \e^{-\ii \Lr*{kx+(k+\Delta k)y-(k+l)x'-(k+l+\Delta l)y'}} \mathcal{D}_{\epsilon_{k,k+\Delta k}, \epsilon_{k+l,k+l+\Delta l}}\\
	= \sum_{k,\Delta k,l,\Delta l} \e^{-\ii \Lr*{k(x+y-x'-y')+\Delta ky-l(x'+y')-\Delta ly'}} \lr*{\JBM}^{2} \delta((k+l)-k)\delta((k+l+\Delta l)-(k+\Delta k))\\
	= \lr*{\JBM}^{2} \sum_{k} \e^{-\ii k(x+y-x'-y')}\sum_{\Delta k,l,\Delta l} \e^{-\ii \Lr*{\Delta ky-l(x'+y')-\Delta ly'}} \delta(l)\delta(l+\Delta l-\Delta k)\\
	= \lr*{\JBM}^{2} M\delta(x+y-x'-y') \sum_{\Delta k,\Delta l} \e^{-\ii \Lr*{\Delta ky-\Delta ly'}} \delta(\Delta l-\Delta k)\\
	= \lr*{\JBM}^{2} M\delta(x+y-x'-y') \sum_{\Delta k} \e^{\ii\Delta k(y'-y)}\\
	= \lr*{\JBM}^{2} M^{2} \delta(x+y-x'-y') \delta(y-y'),
\end{multline}
where $\delta(x)\equiv\delta_{x,0}$ is the Kronecker delta.
From the analogous treatment in all diagonal blocks, we have
\begin{subequations}
\begin{align}
	\label{eq:g1-rho}\Lr*{G_{\rho\rho}        }_{(x,y), (x',y')}^{(1)} & = 2\JOBM M\delta(x-x'),\\
	\label{eq:g1-eps}\Lr*{G_{\epsilon\epsilon}}_{(x,y), (x',y')}^{(1)}   = \Lr*{G_{\zeta\zeta}      }_{(x,y), (x',y')}^{(1)} & = \lr*{\JBM}^{2} M^{2} \delta(x+y-x'-y') \delta(y-y'),\\
	\label{eq:g1-eta}\Lr*{G_{\eta\eta}        }_{(x,y), (x',y')}^{(1)}   = \Lr*{G_{\xi\xi}          }_{(x,y), (x',y')}^{(1)} & = 2\lr*{\JBM}^{2} M^{2} \delta(x+y-x'-y') \delta(y-y').
\end{align}
\end{subequations}

\subsection{Contributions involving the effective Hamiltonian}
From rule~\eqref{eq:rule-trot-relat}, one can see that the third term in Eq.~\eqref{eq:deri}
will always contain at least the same dependence on Trotter indices as the second term.
Therefore, out of the two, focusing on the third term only captures the dependence of both.
Moreover, out of the two terms in the derivatives of the Hamiltonian~[Eqs.~\eqref{eq:ham-derivs-a}-\eqref{eq:ham-derivs-e}],
considering either one alone is enough to capture the dependence on Trotter indices.
Hence, the dependence can be found from analyzing terms of the form
\begin{equation}
	\avg*{p_{k,\alpha}p_{k+\Delta k,\beta}p_{k+l,\alpha}p_{k+l+\Delta l,\beta}}
\end{equation}
or similar but with fewer $p$ variables under the average
when one or both derivatives are over $\rho$.
The exact dependence on Trotter indices varies among blocks of $G^{(3)}$ contributions.

\subsubsection{Example: $\partial^{2} \mathcal{F} / \partial \epsilon \partial \epsilon$}
The block dependent on most Trotter indices is
\begin{equation}
	\Lr*{G_{\epsilon\epsilon}}_{(k, k+\Delta k), (k+l, k+l+\Delta l)}^{(3)}
	= \avg*{\frac{\partial(-\beta \mathcal{H})}{\partial \epsilon_{\alpha, k, k+\Delta k}}\frac{\partial(-\beta \mathcal{H})}{\partial \epsilon_{\alpha, k+l, k+l+\Delta l}}}
	\sim \avg*{p_{k}p_{k+\Delta k}p_{k+l}p_{k+l+\Delta l}}
	\equiv \mathscr{f}(\Delta k, l, \Delta l),
\end{equation}
where by $\sim$ we denote the extraction of a single term
that is sufficient for checking the dependence on Trotter indices.
Upon Fourier transformation it becomes
\begin{multline}
	\Lr*{G_{\epsilon\epsilon}}_{(x,y), (x',y')}^{(3)}
	\sim \sum_{k,\Delta k, l, \Delta l} \e^{-\ii \Lr*{kx+(k+\Delta k)y-(k+l)x'-(k+l+\Delta l)y'}} \mathscr{f}(\Delta k, l, \Delta l)\\
	= \sum_{k}\e^{-\ii k(x+y-x'-y')}\sum_{\Delta k, l, \Delta l} \e^{-\ii \Lr*{\Delta k y-l(x'+y')-\Delta l y'}} \mathscr{f}(\Delta k, l, \Delta l)\\
	= \delta(x+y-x'-y') \sum_{\Delta k, l, \Delta l} \e^{-\ii \Lr*{\Delta k y-l(x'+y')-\Delta l y'}} \mathscr{f}(\Delta k, l, \Delta l).
\end{multline}
By choosing $s=x+y$ and $s'=x'+y'$ and relabeling the matrix indices, we obtain
\begin{equation}
	\Lr*{G_{\epsilon\epsilon}}_{(x,y), (x',y')}^{(3)} = \Lr*{G_{\epsilon\epsilon}}_{(s,y), (s',y')}^{(3)}
	= \delta(s-s') \sum_{\Delta k, l, \Delta l} \e^{\ii \lr*{ls - \Delta k y + \Delta l y'}} \Lr*{G_{\epsilon\epsilon}}_{\lr{k,k+\Delta k}, \lr{k+l,k+l+\Delta l}}^{(3)},
\end{equation}
which is a block-diagonal matrix with $M$ independent blocks labeled by $s$.
Note that the final expression still contains the index $k$
even though there is no longer a summation over it.
We keep it for notation consistency, while it can be picked arbitrarily.

\subsubsection{Example: $\partial^{2} \mathcal{F} / \partial \eta \partial \rho$}
Another illustrious example is
\begin{equation}
	\Lr*{G_{\eta\rho}}_{(k, k+\Delta k), (k+l)}^{(3)}
	= \avg*{\frac{\partial(-\beta \mathcal{H})}{\partial \eta_{\alpha,\alpha', k, k+\Delta k}}\frac{\partial(-\beta \mathcal{H})}{\partial \rho_{\alpha, k+l}}}
	\sim \avg*{p_{k}p_{k+l}}\avg*{p_{k+\Delta k}}
	\equiv \mathscr{f}(l).
\end{equation}
Fourier transformation analogous to the one from the previous section yields
\begin{equation}
	\Lr*{G_{\eta\rho}}_{(x,y), (x')}^{(3)}
	\sim \sum_{k,\Delta k, l} \e^{-\ii \Lr*{kx+(k+\Delta k)y-(k+l)x'}} \mathscr{f}(l)
	= \sum_{k}\e^{-\ii k(x+y-x')}\sum_{\Delta k} \e^{-\ii \Delta k y}\sum_{l} \e^{\ii lx'} \mathscr{f}(l).
\end{equation}
For consistency with the previous case, we choose $s=x+y$ and $s'=x'$ and arrive at
\begin{equation}
	\Lr*{G_{\eta\rho}}_{(s,y), (s')}^{(3)} = \delta(s-s')\delta(y)\sum_{l} \e^{\ii ls} \Lr*{G_{\eta\rho}}_{\lr{k,k+\Delta k}, \lr{k+l}}^{(3)}.
\end{equation}

\subsubsection{All matrix elements}
After working out all unique kinds of second derivatives in the same manner, we obtain
\begin{subequations}
\begin{align}
	\label{eq:g3-rr}\Lr*{G_{\rho\rho        }}_{(s  ),(s'   )}^{(3)} &= \delta(s-s')\sum_{l}                     \e^{\ii ls}                            \Lr*{G_{\rho\rho        }}_{\lr{k}, \lr{k+l}}^{(3)},  \\
	\label{eq:g3-er}\Lr*{G_{\epsilon\rho    }}_{(s,y),(s'   )}^{(3)} &= \delta(s-s')\sum_{\Delta k,l}            \e^{\ii \lr*{ls-\Delta ky}}            \Lr*{G_{\epsilon\rho    }}_{\lr{k,k+\Delta k}, \lr{k+l}}^{(3)},  \\
	\label{eq:g3-nr}\Lr*{G_{\eta\rho        }}_{(s,y),(s'   )}^{(3)} &= \delta(s-s')\delta(y) \sum_{l}           \e^{\ii ls}                            \Lr*{G_{\eta\rho        }}_{\lr{k,k+\Delta k}, \lr{k+l}}^{(3)},  \\
	\label{eq:g3-ee}\Lr*{G_{\epsilon\epsilon}}_{(s,y),(s',y')}^{(3)} &= \delta(s-s')\sum_{\Delta k, l, \Delta l} \e^{\ii \lr*{ls+\Delta ly'-\Delta ky}} \Lr*{G_{\epsilon\epsilon}}_{\lr{k,k+\Delta k}, \lr{k+l,k+l+\Delta l}}^{(3)},  \\
	\label{eq:g3-ne}\Lr*{G_{\eta\epsilon    }}_{(s,y),(s',y')}^{(3)} &= \delta(s-s')\delta(y) \sum_{l, \Delta l} \e^{\ii \lr*{ls-\Delta ly'}}           \Lr*{G_{\eta\epsilon    }}_{\lr{k,k+\Delta k}, \lr{k+l,k+l+\Delta l}}^{(3)},  \\
	\label{eq:g3-nn}\Lr*{G_{\eta\eta        }}_{(s,y),(s',y')}^{(3)} &= \delta(s-s')\delta(y-y') \sum_{l, d}     \e^{\ii \lr*{ls+dy-ly'}}               \Lr*{G_{\eta\eta        }}_{\lr{k,k+\Delta k}, \lr{k+l,k+l+\Delta l}}^{(3)},
\end{align}
\end{subequations}
where in the last equation $d = l+\Delta l-\Delta k$.
The elements in other blocks may be found by substituting $\epsilon\to\zeta$ or $\eta\to\xi$,
while the elements with swapped block indices (i.e., of the transposed blocks) may be obtained 
by swapping $y\leftrightarrow -y'$ and $\Delta k\leftrightarrow \Delta l$.

\subsection{Decoupling of the matrix}
\begin{figure}[tb]
	\includegraphics{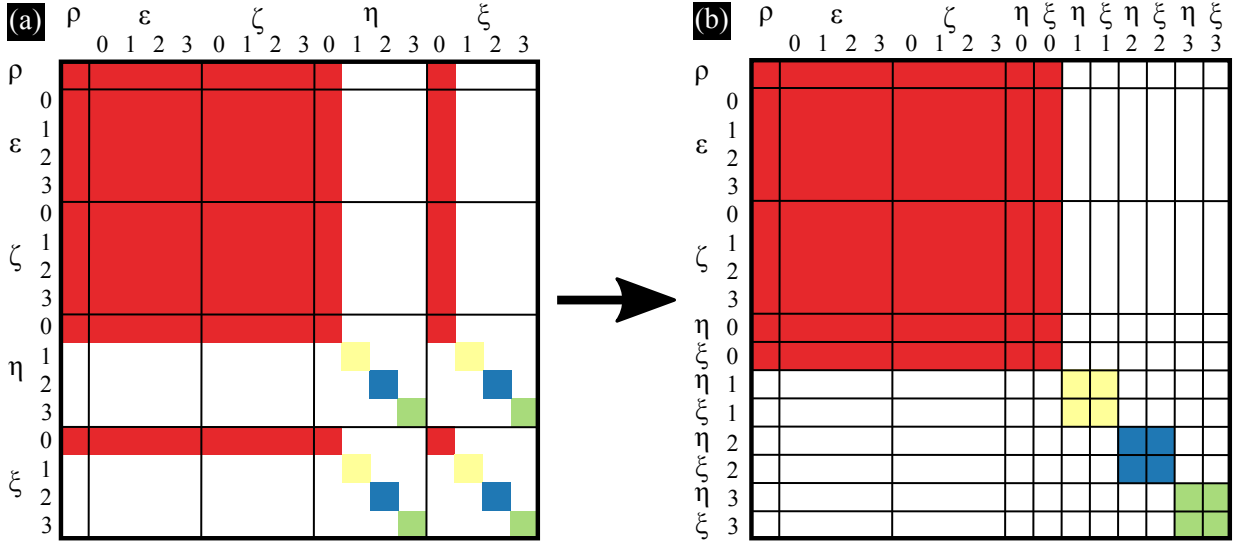}
	\caption{
		(a) Pictorial representation of one of the constant-$s$ blocks of the matrix $G$,
		showing the decoupling in the Fourier space due to invariances in the Trotter dimensions,
		reflected also in Kronecker deltas in Eqs.~\eqref{eq:g3-rr}-\eqref{eq:g3-nn}.
		Vanishing elements are white,
		while each of the colors marks a set of elements decoupled from other colors.
		Subsets of indices corresponding to various variables
		are marked by Greek letters and separated by lines.
		Values of the Fourier-transformed Trotter indices $y$
		are marked by numbers 0-3 (the picture is drawn for $M=4$).
		Each of the marked blocks is further indexed by replica indices,
		which, however, do not play any role in the current decoupling.
		(b) The same matrix after rearrangement of indices,
		showing the decoupling into block-diagonal matrix.
	}
	\label{fig:matrix-nonzero}
\end{figure}
Putting together $G^{(1)}$ and $G^{(3)}$,
we recover the full dependence of $\Lr*{G_{uv}}_{(s,y), (s',y')}$ on Trotter indices.
We recognize the expressions $\delta(x-x')$ in Eq.~\eqref{eq:g1-rho}
and $\delta(x+y-x'-y')$ in Eqs.~\eqref{eq:g1-eps}-\eqref{eq:g1-eta}
as $\delta(s-s')$ from Eqs.~\eqref{eq:g3-rr}-\eqref{eq:g3-nn}.
With this, we notice that non-zero elements allowed by $G^{(1)}$
are a subset of those allowed by $G^{(3)}$.
Hence, the dependence on Trotter indices can be read directly from Eqs.~\eqref{eq:g3-rr}-\eqref{eq:g3-nn}.

First of all, only the elements diagonal in $s$ are non-zero,
which is what we expected from Fourier transforming a translationally invariant expression.
However, there is additional decoupling present.
Only $y=0$, $y'=0$, or $y=y'$ elements remain in some of the $uv$ blocks,
splitting the $s$-blocks further.
In Fig.~\ref{fig:matrix-nonzero} we graphically illustrate the non-zero elements
allowed within a single $s$-block.
In total, the initial matrix $G$ decouples into:
\begin{itemize}
\item $M$ matrices $g^{(s)}$, for each $s\in\LR*{0,\ldots,M-1}$,
composed of the whole block spanned by $\rho$, $\epsilon$, $\zeta$ indices,
the blocks spanned by those of $\eta$, $\xi$ indices where $y=y'=0$,
and the off-diagonal blocks coupling them
[columns (rows) from the $\LR{\rho, \epsilon, \zeta}$-$\LR{\eta, \xi}$ ($\LR{\eta, \xi}$-$\LR{\rho, \epsilon, \zeta}$) block with $y=0$ ($y'=0$)].
\item $M(M-1)$ matrices $g^{(\tilde{s},\tilde{y})}$, for each $\tilde{s}\in\LR*{0,\ldots,M-1}$, $\tilde{y}\in\LR*{1,\ldots,M-1}$,
composed of blocks spanned by those of $\eta$, $\xi$ indices where $s=\tilde{s}$, $y=\tilde{y}$,
and the off-diagonal blocks coupling them.
\end{itemize}

\section{Hessian matrix elements}\label{sec:matrixel}
Once we have simplified the Trotter indices, we invert the order of index nesting
with respect to what we used in the previous section.
Now, the notation is
\begin{equation}
	\Lr*{\Lr*{G_{uv}}_{\LR{\alpha},\LR{\beta}}}_{y,y'},
\end{equation}
where the meaning of $u$, $v$ and $\LR*{\alpha}, \LR*{\beta}$ is the same as it was in Eq.~\eqref{eq:struct-ka},
while $y$ and $y'$ are transformed Trotter indices.
Throughout most of this section, we focus on one of the multiple analogous matrices $g^{(s)}$.
We drop the $s$ index, but implicitly work in a case of constant value of $s$.
In the analytical treatment of a single matrix $g$,
we are no longer interested in the dependence on indices in the reciprocal Trotter space.
In the following derivations, we drop the indices $y$ and  $y'$,
keeping in mind that the dimensionality of $\Lr*{G_{uv}}_{\LR{\alpha},\LR{\beta}}$
may vary depending on $u$ and $v$.
Nevertheless, we will bring up these indices where needed.

We name the blocks of $g$ according to:
\begin{equation}\label{eq:mat-blocks}
	g = ~
	\begin{blockarray}{cccccc}
		\rho_{\alpha} & \epsilon_{\alpha} & \zeta_{\alpha} & \eta_{\alpha\alpha'} & \xi_{\alpha\alpha'}\\[4pt]
		\begin{block}{(ccccc)l}
			\mathbfbb{A} & \mathbfbb{B} & \mathbfbb{C} & \mathbfbb{D} & \mathbfbb{E} & ~~\rho_{\alpha}\\
			\mathbfbb{B} & \mathbfbb{P} & \mathbfbb{Q} & \mathbfbb{S} & \mathbfbb{T} & ~~\epsilon_{\alpha}\\
			\mathbfbb{C} & \mathbfbb{Q} & \mathbfbb{R} & \mathbfbb{U} & \mathbfbb{V} & ~~\zeta_{\alpha}\\
			\mathbfbb{D} & \mathbfbb{S} & \mathbfbb{U} & \mathbfbb{X} & \mathbfbb{Y} & ~~\eta_{\alpha\alpha'}\\
			\mathbfbb{E} & \mathbfbb{T} & \mathbfbb{V} & \mathbfbb{Y} & \mathbfbb{Z} & ~~\xi_{\alpha\alpha'}\\
		\end{block}
	\end{blockarray}.
\end{equation}
There, each block contains only a few distinct kinds of matrix elements,
since the value of the second derivative of $\mathcal{F}$ [Eq.~\eqref{eq:deri}] depends only on mutual equalities between the replicas involved
and not on the specific values of the replica indices.
For example, block $\mathbfbb{A}$ contains two types of elements:
\begin{subequations}
\begin{align}
	\mathbfbb{A}_{\alpha\alpha} &\equiv \mathbb{A}' = \Lr*{G_{\rho\rho}}_{\lr{\alpha},\lr{\alpha}},\\
	\mathbfbb{A}_{\alpha\beta } &\equiv \mathbb{A}  = \Lr*{G_{\rho\rho}}_{\lr{\alpha},\lr{\beta}}.
\end{align}
\end{subequations}
In the case of blocks $\mathbfbb{P}, \mathbfbb{Q}$ and $\mathbfbb{R}$,
the replica-indexed elements are matrices themselves,
indexed by Fourier-transformed Trotter indices,
for example,
\begin{subequations}
\begin{align}
	\mathbfbb{P}_{\alpha\alpha} &\equiv \mathbb{P}',\quad \mathbb{P}'_{yy'} = \Lr*{\Lr*{G_{\epsilon\epsilon}}_{\lr{\alpha},\lr{\alpha}}}_{(y),(y')},\\
	\mathbfbb{P}_{\alpha\beta } &\equiv \mathbb{P} ,\quad \mathbb{P}_{yy'}  = \Lr*{\Lr*{G_{\epsilon\epsilon}}_{\lr{\alpha},\lr{\beta}}}_{(y),(y')}.
\end{align}
\end{subequations}
The blocks indexed by two replica indices ($\mathbfbb{X}$, $\mathbfbb{Y}$, and $\mathbfbb{Z}$)
contain three distinct elements, with two, one, or zero mutual replica indices:
\begin{subequations}
\begin{align}
	\mathbfbb{Y}_{(\alpha\beta) (\alpha\beta) } &\equiv \mathbb{Y}'' = \Lr*{G_{\eta\xi}}_{\lr{\alpha\beta},\lr{\alpha\beta}},\\
	\mathbfbb{Y}_{(\alpha\beta) (\alpha\gamma)} &\equiv \mathbb{Y}'  = \Lr*{G_{\eta\xi}}_{\lr{\alpha\beta},\lr{\alpha\gamma}},\\
	\mathbfbb{Y}_{(\alpha\beta) (\gamma\delta)} &\equiv \mathbb{Y}   = \Lr*{G_{\eta\xi}}_{\lr{\alpha\beta},\lr{\gamma\delta}}.
\end{align}
\end{subequations}
The elements of remaining blocks are defined in the same way as above,
with the number of primes ($'$) equal to the number of mutual replicas.

The decoupled matrices $g^{(s,y)}$ contain blocks of the same type as the main $g^{(s)}$ ones
but with a non-zero value of the label $y$, i.e.,
\begin{equation}
	g^{(s,y)} = ~
	\begin{blockarray}{ccc}
		\eta_{\alpha\alpha'} & \xi_{\alpha\alpha'}\\[4pt]
		\begin{block}{(cc)l}
			\mathbfbb{X}^{(y)} & \mathbfbb{Y}^{(y)} & ~~\eta_{\alpha\alpha'}\\
			\mathbfbb{Y}^{(y)} & \mathbfbb{Z}^{(y)} & ~~\xi_{\alpha\alpha'}\\
		\end{block}
	\end{blockarray},
\end{equation}
where, e.g.,
\begin{equation}
	\mathbfbb{Y}^{(y)}_{(\alpha\beta) (\alpha\beta) } \equiv \mathbb{Y}''_{(y)} = \Lr*{\Lr*{G_{\eta\xi}}_{\lr{\alpha\beta},\lr{\alpha\beta}}}_{(y),(y)}.
\end{equation}

\section{Eigenvectors and eigenvalues}\label{sec:eigenvectors}
The matrices we wish to diagonalize do not have exactly the same structure as the AT matrix.
However, we encounter the same blocks as in the latter.
The difference is that we have multiple blocks of the same kind.
We expect that expanding the postulated eigenvectors to contain multiple elements of the same kind
but keeping the premise of a vector being symmetric under interchange of all but $0/1/2$ indices
should yield correct eigenvectors of our matrices.
Thus, we postulate several kinds of eigenvectors
and find corresponding eigenvalues.
Similarly to Eq.~\eqref{eq:mat-blocks}, we name the blocks of these vectors according to
\begin{equation}
	\begin{array}{rcccccl}
		             & \rho_{\alpha} & \epsilon_{\alpha} & \zeta_{\alpha} & \eta_{\alpha\alpha'} & \xi_{\alpha\alpha'}\\
		\mu = (& \vecel{a}     & \vecel{b}         & \vecel{c}      & \vecel{d}            & \vecel{e} & )^\mathrm{T}
	\end{array}
\end{equation}

\subsection{Trial eigenvectors}
The first type of eigenvector, $\mm{0}{}$, is symmetric in replica space,
\begin{equation}
	\vecel{a}_{\alpha}        = a,\qquad
	\vecel{b}_{\alpha}        = b,\qquad
	\vecel{c}_{\alpha}        = c,\qquad
	\vecel{d}_{\alpha\alpha'} = d,\qquad
	\vecel{e}_{\alpha\alpha'} = e,
\end{equation}
where $a,d,e$ are numbers, while $b,c$ are Trotter-indexed vectors.
We multiply $g$ by it and check the eigenproblem equation
\begin{equation}
	g\mm{0}{} = \lambda \mm{0}{}.
\end{equation}
For example, in the first block, we obtain
\begin{multline}
	\lambda a = \Lr*{\Lr[\big]{g\mm{0}{}}_{\rho}}_{\alpha} =
	  \mathbb{A'}a + (n-1)\mathbb{A}a
	+ \mathbb{B'}b + (n-1)\mathbb{B}b
	+ \mathbb{C'}c + (n-1)\mathbb{C}c\\
	+ (n-1)\mathbb{D'}d + \binom{n}{2}\mathbb{D}d
	+ (n-1)\mathbb{E'}e + \binom{n}{2}\mathbb{E}e,
\end{multline}
where $\binom{n}{k}$ is the standard binomial coefficient.
Equations for all blocks combined give an effective eigenproblem
\begin{equation}
	g_{0} (a, b, c, d, e)^{\mathrm{T}} = \lambda (a, b, c, d, e)^{\mathrm{T}},
\end{equation}
with
\begin{multline}\label{eq:effmat-0}
	g_{0} = \\ \footnotesize	
	\begin{pmatrix}
		 \mathbb{A'}+\nminus{1}\mathbb{A} &  \mathbb{B'}+\nminus{1}\mathbb{B} &  \mathbb{C'}+\nminus{1}\mathbb{C} & \nminus{1}\mathbb{D'}+\binom{n-1}{2}\mathbb{D}               & \nminus{1}\mathbb{E'}+\binom{n-1}{2}\mathbb{E}              \\
		 \mathbb{B'}+\nminus{1}\mathbb{B} &  \mathbb{P'}+\nminus{1}\mathbb{P} &  \mathbb{Q'}+\nminus{1}\mathbb{Q} & \nminus{1}\mathbb{S'}+\binom{n-1}{2}\mathbb{S}               & \nminus{1}\mathbb{T'}+\binom{n-1}{2}\mathbb{T}              \\
		 \mathbb{C'}+\nminus{1}\mathbb{C} &  \mathbb{Q'}+\nminus{1}\mathbb{Q} &  \mathbb{R'}+\nminus{1}\mathbb{R} & \nminus{1}\mathbb{U'}+\binom{n-1}{2}\mathbb{U}               & \nminus{1}\mathbb{V'}+\binom{n-1}{2}\mathbb{V}              \\
		2\mathbb{D'}+\nminus{2}\mathbb{D} & 2\mathbb{S'}+\nminus{2}\mathbb{S} & 2\mathbb{U'}+\nminus{2}\mathbb{U} & \mathbb{X''}+2\nminus{2}\mathbb{X'}+\binom{n-2}{2}\mathbb{X} & \mathbb{Y''}+2\nminus{2}\mathbb{Y'}+\binom{n-2}{2}\mathbb{Y}\\
		2\mathbb{E'}+\nminus{2}\mathbb{E} & 2\mathbb{T'}+\nminus{2}\mathbb{T} & 2\mathbb{V'}+\nminus{2}\mathbb{V} & \mathbb{Y''}+2\nminus{2}\mathbb{Y'}+\binom{n-2}{2}\mathbb{Y} & \mathbb{Z''}+2\nminus{2}\mathbb{Z'}+\binom{n-2}{2}\mathbb{Z}\\
	\end{pmatrix},
\end{multline}
which is now a fixed-size eigenproblem
instead of the replica-dependent size of the original matrix.
Hence we can calculate its eigenvalues numerically
after taking the $n\to 0$ limit.

The second type of eigenvectors, $\mu^{(\theta)}$,
is symmetric under interchange of all but one selected replica index $\theta$.
Its elements are defined as
\begin{equation}
	\vecel{a}_{\alpha} = \begin{cases}
		a', & \mbox{ when } \alpha = \theta\\
		a , & \mbox{ when } \alpha \neq \theta
	\end{cases},\qquad\qquad
	\vecel{d}_{\alpha\alpha'} = \begin{cases}
		d', & \mbox{ when } (\alpha = \theta) \mbox{ or } (\alpha' = \theta)\\
		d, & \mbox{ when } (\alpha \neq \theta) \mbox{ and } (\alpha' \neq \theta)
	\end{cases}.
\end{equation}
Elements of blocks $\vecel{b}$ and $\vecel{c}$ are defined analogously as those of $\vecel{a}$,
while elements of block $\vecel{e}$ are analogous to those of $\vecel{d}$.
The requirement of orthogonality to $\mm{0}{}$ links the variables in each pair,
\begin{align}
	a = -(n-1)a',        \quad
	b = -(n-1)b',        \quad
	c = -(n-1)c',        \quad
	d = -\frac{n-2}{2}d',\quad
	e = -\frac{n-2}{2}e',
\end{align}
which leaves a total of 5 independent variables, yielding the effective eigenproblem matrix
\begin{multline}\label{eq:effmat-1}
	g_{1} =\\ \footnotesize
	\begin{pmatrix}
		      \mathbb{A'}-\mathbb{A}  &       \mathbb{B'}-\mathbb{B}  &       \mathbb{C'}-\mathbb{C}  & \mathbb{D}-\mathbb{D'}                        & \mathbb{E}-\mathbb{E'}                       \\
		      \mathbb{B'}-\mathbb{B}  &       \mathbb{P'}-\mathbb{P}  &       \mathbb{Q'}-\mathbb{Q}  & \mathbb{S}-\mathbb{S'}                        & \mathbb{T}-\mathbb{T'}                       \\
		      \mathbb{C'}-\mathbb{C}  &       \mathbb{Q'}-\mathbb{Q}  &       \mathbb{R'}-\mathbb{R}  & \mathbb{U}-\mathbb{U'}                        & \mathbb{V}-\mathbb{V'}                       \\
		\nminus{2}(\mathbb{D}-\mathbb{D'}) & \nminus{2}(\mathbb{S}-\mathbb{S'}) & \nminus{2}(\mathbb{U}-\mathbb{U'}) & \mathbb{X''}+\nminus{4}\mathbb{X'}-\nminus{3}\mathbb{X} & \mathbb{Y''}+\nminus{4}\mathbb{Y'}-\nminus{3}\mathbb{Y}\\
		\nminus{2}(\mathbb{E}-\mathbb{E'}) & \nminus{2}(\mathbb{T}-\mathbb{T'}) & \nminus{2}(\mathbb{V}-\mathbb{V'}) & \mathbb{Y''}+\nminus{4}\mathbb{Y'}-\nminus{3}\mathbb{Y} & \mathbb{Z''}+\nminus{4}\mathbb{Z'}-\nminus{3}\mathbb{Z}\\
	\end{pmatrix}.
\end{multline}

The third type of eigenvectors, $\mu^{(\theta, \theta')}$,
is symmetric under interchange of all but two replica indices ($\theta, \theta'$).
Its elements are defined as
\begin{subequations}
\begin{align}
	\vecel{a}_{\alpha} &= \begin{cases}
		a', & \mbox{ when } (\alpha = \theta) \mbox{ or } (\alpha = \theta')\\
		a,  & \mbox{ when } (\alpha \neq \theta) \mbox{ and } (\alpha \neq \theta')
	\end{cases},\\
	\vecel{d}_{\alpha\alpha'} &= \begin{cases}
		d'', & \mbox{ when } (\alpha,\alpha') = (\theta,\theta')\\
		d',  & \mbox{ when } (\alpha,\alpha') \mbox { and } (\theta,\theta') \mbox{ have one common element}\\
		d,   & \mbox{ when } (\alpha,\alpha') \mbox { and } (\theta,\theta') \mbox{ have no common elements}
	\end{cases}.
\end{align}
\end{subequations}
Again, elements of blocks $\vecel{b}$ and $\vecel{c}$ are defined analogously as those of $\vecel{a}$,
while elements of block $\vecel{e}$ are analogous to those of $\vecel{d}$.
Orthogonalization to both previous types of eigenvectors yields
\begin{equation}
	a = a' = 0,\qquad
	b = b' = 0,\qquad
	c = c' = 0,
\end{equation}
and
\begin{equation}
	d = \binom{n-2}{2}d'',\quad d' = -\frac{n-3}{2}d'',\qquad
	e = \binom{n-2}{2}e'',\quad e' = -\frac{n-3}{2}e'',
\end{equation}
leaving only 2 independent variables $d''$ and $e''$.
The associated matrix is
\begin{equation}\label{eq:effmat-2}
	g_{2} = \mbox{$\footnotesize
	\begin{pmatrix}
		\mathbb{X''}-2\mathbb{X'}+\mathbb{X} & \mathbb{Y''}-2\mathbb{Y'}+\mathbb{Y}\\
		\mathbb{Y''}-2\mathbb{Y'}+\mathbb{Y} & \mathbb{Z''}-2\mathbb{Z'}+\mathbb{Z}\\
	\end{pmatrix}$}.
\end{equation}
It should be kept in mind that $g_{2}$ (and its elements) implicitly depends on the reciprocal Trotter index $s$.
Where the latter is important, we refer to the matrix as $g_{2}^{(s)}$.

\subsection{Effective eigenproblems}
We notice that in the limit of $n\to 0$, both $g_{0}$ and $g_{1}$
reduce to the same matrix of the form
\begin{equation}\label{eq:effmat-01}
	g_{01} = \mbox{$\footnotesize
	\begin{pmatrix}
		  \mathbb{A'}-\mathbb{A}  &   \mathbb{B'}-\mathbb{B}  &   \mathbb{C'}-\mathbb{C}  & \mathbb{D}-\mathbb{D'}                & \mathbb{E}-\mathbb{E'}               \\
		  \mathbb{B'}-\mathbb{B}  &   \mathbb{P'}-\mathbb{P}  &   \mathbb{Q'}-\mathbb{Q}  & \mathbb{S}-\mathbb{S'}                & \mathbb{T}-\mathbb{T'}               \\
		  \mathbb{C'}-\mathbb{C}  &   \mathbb{Q'}-\mathbb{Q}  &   \mathbb{R'}-\mathbb{R}  & \mathbb{U}-\mathbb{U'}                & \mathbb{V}-\mathbb{V'}               \\
		2(\mathbb{D'}-\mathbb{D}) & 2(\mathbb{S'}-\mathbb{S}) & 2(\mathbb{U'}-\mathbb{U}) & \mathbb{X''}-4\mathbb{X'}+3\mathbb{X} & \mathbb{Y''}-4\mathbb{Y'}+3\mathbb{Y}\\
		2(\mathbb{E'}-\mathbb{E}) & 2(\mathbb{T'}-\mathbb{T}) & 2(\mathbb{V'}-\mathbb{V}) & \mathbb{Y''}-4\mathbb{Y'}+3\mathbb{Y} & \mathbb{Z''}-4\mathbb{Z'}+3\mathbb{Z}\\
	\end{pmatrix}$},
\end{equation}
while $g_{2}$ does not depend on $n$
and thus stays unchanged.
In the case of AT, the matrix $g_{2}$ was a single element only,
and it was the only eigenvalue of $g$ that could have negative values.

The first kind of eigenvectors produces $2M+3$ eigenvalues, each with the multiplicity of 1.
The second kind produces another $2M+3$ eigenvalues, each with the multiplicity of $n-1$.
Finally, the third kind of eigenvectors produces two eigenvalues with the multiplicity of $n(n-3)/2$.
The total number of eigenvalues found this way is $n^2+2Mn$,
equal to the dimension of the matrix $g^{(s)}$.
Thus, all the eigenvalues of each of the $g^{(s)}$ matrices
are among the eigenvalues of corresponding matrices $g_{01}^{(s)}$ and $g_{2}^{(s)}$.

The matrices $g^{(s,y)}$ are equivalent to the bottom right corner
of matrices $g^{(s)}$ (but with matrix elements defined with $y\neq 0$).
Thus, their eigenvalues are analogous as well.
The effective eigenproblems to solve in this case are defined by matrices
\begin{equation}\label{eq:effmat-01-sy}
	g_{01}^{(s,y)} = \mbox{$\footnotesize
	\begin{pmatrix}
		\mathbb{X''}^{(y)}-4\mathbb{X'}^{(y)}+3\mathbb{X}^{(y)} & \mathbb{Y''}^{(y)}-4\mathbb{Y'}^{(y)}+3\mathbb{Y}^{(y)}\\
		\mathbb{Y''}^{(y)}-4\mathbb{Y'}^{(y)}+3\mathbb{Y}^{(y)} & \mathbb{Z''}^{(y)}-4\mathbb{Z'}^{(y)}+3\mathbb{Z}^{(y)}\\
	\end{pmatrix}$}
\end{equation}
and
\begin{equation}\label{eq:effmat-2-sy}
	g_{2}^{(s,y)} = \mbox{$\footnotesize
	\begin{pmatrix}
		\mathbb{X''}^{(y)}-2\mathbb{X'}^{(y)}+\mathbb{X}^{(y)} & \mathbb{Y''}^{(y)}-2\mathbb{Y'}^{(y)}+\mathbb{Y}^{(y)}\\
		\mathbb{Y''}^{(y)}-2\mathbb{Y'}^{(y)}+\mathbb{Y}^{(y)} & \mathbb{Z''}^{(y)}-2\mathbb{Z'}^{(y)}+\mathbb{Z}^{(y)}\\
	\end{pmatrix}$}.
\end{equation}

All the eigenvalues of the original matrix $G$
are eigenvalues of one of the matrices
$g_{2}^{(s)}$, $g_{01}^{(s)}$, $g_{2}^{(s,y)}$, or $g_{01}^{(s,y)}$ for some $s$ and $y$.
If there is a negative one among them, the solution is unstable.
All these matrices can be evaluated and diagonalized after taking the $n\to 0$ limit,
which makes it possible to be done numerically.

\section{Numerical evaluation}\label{sec:numerical}
We evaluate matrix elements numerically
and use them to construct matrices $g_{2}^{(s)}$, $g_{01}^{(s)}$, $g_{2}^{(s,y)}$, or $g_{01}^{(s,y)}$.
Next, we diagonalize these matrices and analyze the resulting eigenvalues,
looking for the negative ones, which are a fingerprint of instabilities.

For the analysis of the eigenvalues,
we choose a $\mu/U-J/U$ plane at constant $T/U=0.08$ and $J_{0}/U=0.122$,
as it contains all three phases that we can distinguish based on Landau's theory of phase transitions.
In this plane, we choose three cuts:
(A) along varying $J/U$ at constant $\mu/U=0.8$, (B) along varying $J/U$ at constant $\mu/U=0.4$,
and (C) along varying $\mu/U$ at constant $J/U=0.1$.

\begin{figure}[tb]
	\includegraphics[width=0.45\columnwidth]{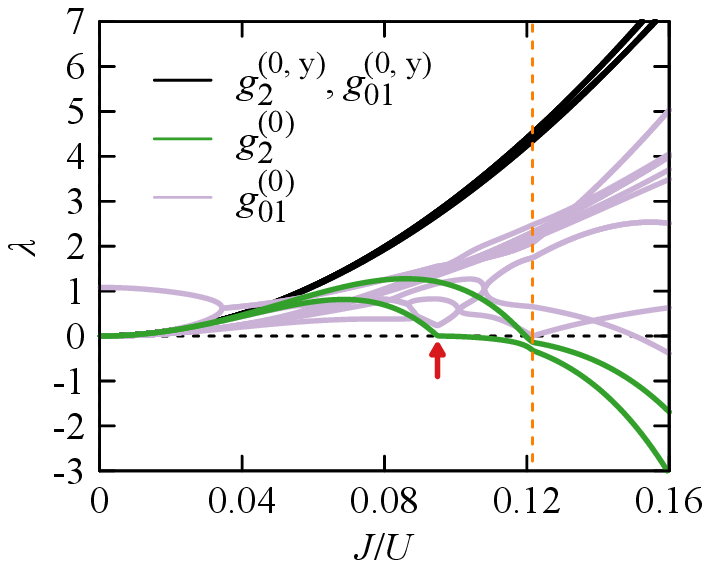}
	\includegraphics[width=0.45\columnwidth]{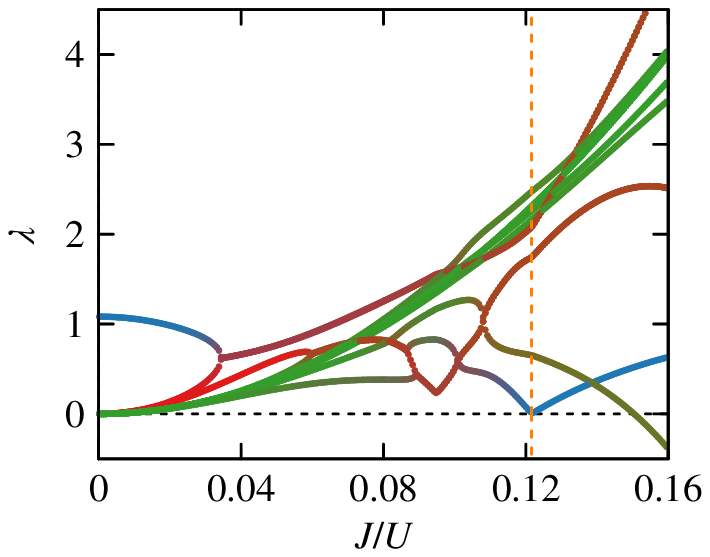}
	\caption{
		(left) Spectra of the matrices $g_{2}^{(0)}$, $g_{01}^{(0)}$, $g_{2}^{(0,y)}$,
		and $g_{01}^{(0,y)}$ for all valid $y$ as a function of $J/U$,
		obtained for $M = 5$, $T/U = 0.08$, $J_{0}/U = 0.122$, and $\mu/U = 0.8$.
		The vertical dashed line marks the phase transition point
		between the superfluid (to the left of the line) and the glass (to the right).
		The red arrow marks the point where a negative eigenvalue appears.
		(right) Spectrum of the matrix $g_{01}^{(0)}$ repeated from panel (a).
		Colors of the lines denote the composition of the eigenvectors associated with the eigenvalues:
		red, green, and blue color components show the size of the eigenvector projection onto subspaces
		spanned by $\lbrace\eta, \xi\rbrace$, $\lbrace\epsilon, \zeta\rbrace$, and $\lbrace\rho\rbrace$, respectively.
	}
	\label{fig:lines-m8}
\end{figure}
We start with the cut (A) where, according to Landau's theory of phase transitions,
there is a phase with $q>0$ and $\varDelta=0$ (glass) above $J/U\approx 0.122$
and a phase with $q>0$ and $\varDelta>0$ (superfluid) below $J/U\approx 0.122$.
In Fig.~\ref{fig:lines-m8}(a), we plot the spectra of matrices
$g_{2}^{(0)}$, $g_{01}^{(0)}$, $g_{2}^{(0,y)}$, and $g_{01}^{(0,y)}$ for all valid $y$.
We find that the instability comes from the matrix $g_{2}^{(0)}$ and occurs above $J/U\approx 0.095$.
This value does not correspond to any of the previously known phase transition points.
Thus, we associate it with the existence of two flavors of a superfluid phase:
a stable one below $J/U\approx 0.095$ and an unstable one between $0.95 \lesssim J/U \lesssim 0.122$.
This finding is consistent with previous works, where a term analogous to this matrix was a fingerprint of the instability.
We notice that all eigenvalues of matrices $g_{2}^{(0,y)}$ and $g_{01}^{(0,y)}$ are very close to $2\lr*{J\beta}^{2}$,
which is the value one would obtain from the bare $\Lr*{G_{\eta\eta}}_{(s,y), (s',y')}^{(1)}$ diagonal matrix element,
without (or with negligibly small) $\Lr*{G_{\eta\eta}}_{(s,y), (s',y')}^{(2)}$ and $\Lr*{G_{\eta\eta}}_{(s,y), (s',y')}^{(3)}$ elements,
as well as without off-diagonal elements.
A possible cause of this behavior might be that $\Lr*{G_{\eta\eta}}_{\lr{k,k+\Delta k}, \lr{k+l,k+l+\Delta l}}^{(3)}$
is only very weakly dependent on Trotter indices,
resulting in its Fourier transform concentrated at $y=0$.
The spectrum of matrix $g_{01}^{(0)}$ is not as simple.
Some of the eigenvalues follow a $\lr*{J\beta}^{2}$ dependence of the $\Lr*{G_{\epsilon\epsilon}}_{(s,y), (s',y')}^{(1)}$ matrix element,
but the behavior of the remaining ones is more complex.
At the superfluid-glass transition, one of the eigenvalues reaches zero, which reflects the marginal stability at the critical point.
Deep inside the unstable (glass) phase, one of them becomes negative.
We analyze the eigenvalues in more detail by looking at the origin of each of them.
In Fig.~\ref{fig:lines-m8}(b), we plot the spectrum of $g_{01}^{(0)}$
and color-code it according to occupations within each block of the matrix.
From this, one can see that the eigenvalues $\approx \lr*{J\beta}^{2}$
are indeed coming from the blocks spanned by $\epsilon$ and $\zeta$ indices.
Apart from these seven uncoupled lines,
all the remaining lines incorporate more than one type of indices.
One can see multiple couplings and anticrossings between these eigenvalues.

\begin{figure}[tb]
	\includegraphics[width=0.45\columnwidth]{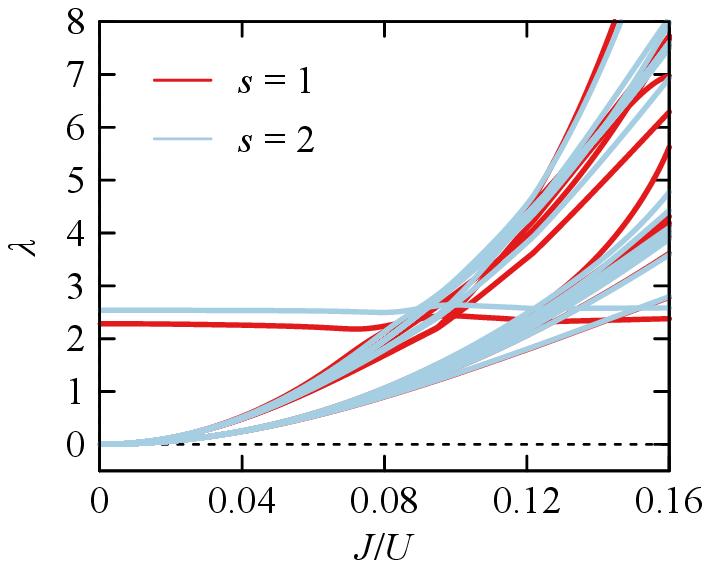}
	\includegraphics[width=0.45\columnwidth]{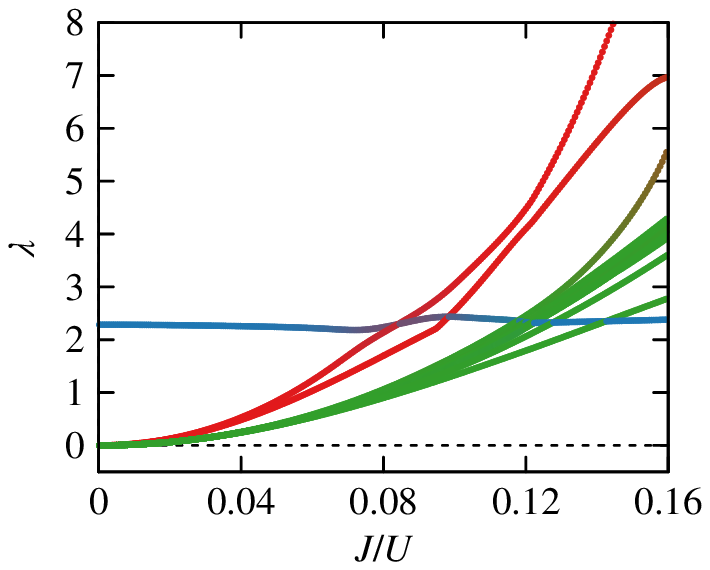}
	\caption{
		(left) Spectra of the matrices $g_{2}^{(s)}$, $g_{01}^{(s)}$, $g_{2}^{(s,y)}$,
		and $g_{01}^{(s,y)}$ for $s=1$, 2 and all valid $y$ as a function of $J/U$,
		obtained for $M = 5$, $T/U = 0.08$, $J_{0}/U = 0.122$, and $\mu/U = 0.8$.
		(right) Spectrum of the matrix $g_{01}^{(1)}$ repeated from panel (a).
		Colors of the lines denote the composition of the eigenvectors associated with the eigenvalues:
		red, green, and blue color components show the size of the eigenvector projection onto subspaces
		spanned by $\lbrace\eta, \xi\rbrace$, $\lbrace\epsilon, \zeta\rbrace$, and $\lbrace\rho\rbrace$, respectively.
	}
	\label{fig:lines-s1}
\end{figure}
Next, in Fig.~\ref{fig:lines-s1}(a), we plot the spectra of matrices
$g_{2}^{(s)}$, $g_{01}^{(s)}$, $g_{2}^{(s,y)}$, and $g_{01}^{(s,y)}$ for $s>0$ and all valid $y$.
These matrices taken at $s$ and $M-s$ are equivalent,
so, since we are working with $M=5$, the only unique cases are $s=1$ and $s=2$.
Again, the eigenvalues of $g_{2}^{(s,y)}$ and $g_{01}^{(s,y)}$ follow $2\lr*{J\beta}^{2}$ dependence.
However, this time the eigenvalues of $g_{2}^{(s)}$ and $g_{01}^{(s)}$ seem to behave trivially as well.
In Fig.~\ref{fig:lines-s1}(b), we plot the latter color-coded in the same manner as we did previously.
We find that there is no coupling between blocks spanned by different indices.
Namely, there are ten eigenvalues $\approx \lr*{J\beta}^{2}$, two $\approx 2\lr*{J\beta}^{2}$,
and one nearly constant (although not equal to the expected value of $2J_{0}\beta$),
with only minimal mixing between blocks.

\begin{figure}[tb]
	\includegraphics[width=0.45\columnwidth]{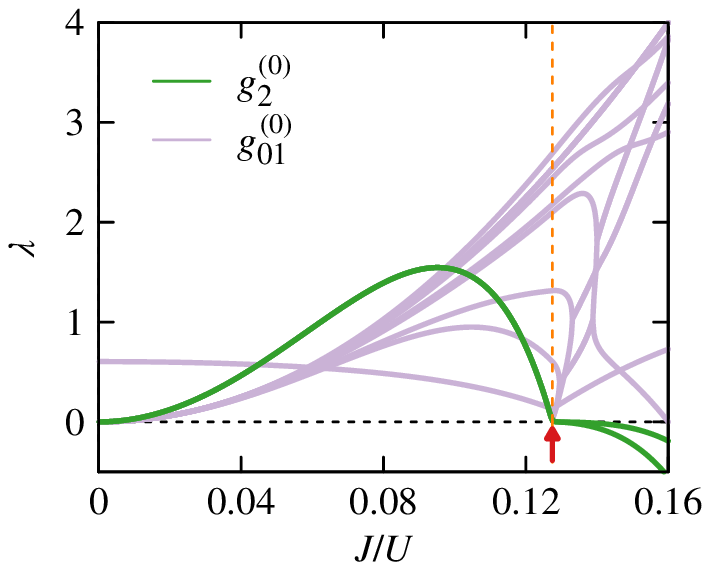}
	\includegraphics[width=0.45\columnwidth]{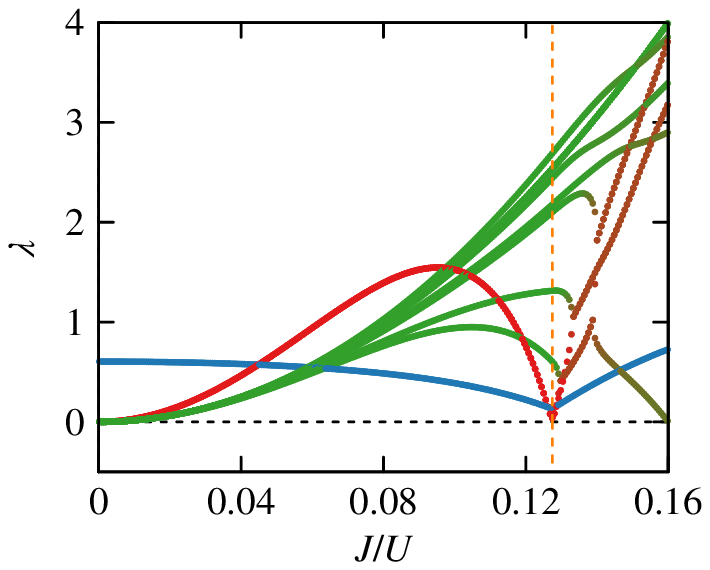}
	\caption{
		(left) Spectra of the matrices $g_{2}^{(0)}$ and $g_{01}^{(0)}$ for all valid $y$ as a function of $J/U$,
		obtained for $M = 5$, $T/U = 0.08$, $J_{0}/U = 0.122$, and $\mu/U = 0.4$.
		The vertical dashed line marks the phase transition point
		between the disordered (to the left of the line) and glass (to the right) phases.
		The red arrow marks the point where a negative eigenvalue appears.
		(right) Spectrum of the matrix $g_{01}^{(0)}$ repeated from panel (a).
		Colors of the lines denote the composition of the eigenvectors associated with the eigenvalues:
		red, green, and blue color components show the size of the eigenvector projection onto subspaces
		spanned by $\lbrace\eta, \xi\rbrace$, $\lbrace\epsilon, \zeta\rbrace$, and $\lbrace\rho\rbrace$, respectively.
	}
	\label{fig:lines-m4}
\end{figure}
\begin{figure}[tb]
	\includegraphics[width=0.45\columnwidth]{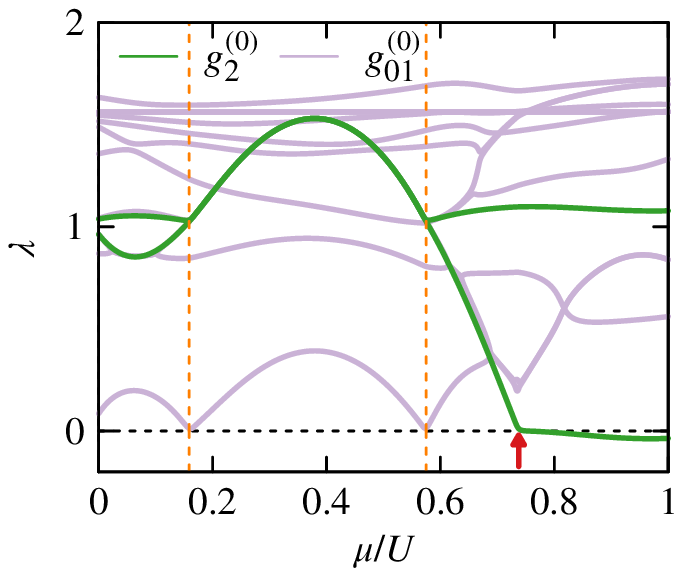}
	\includegraphics[width=0.45\columnwidth]{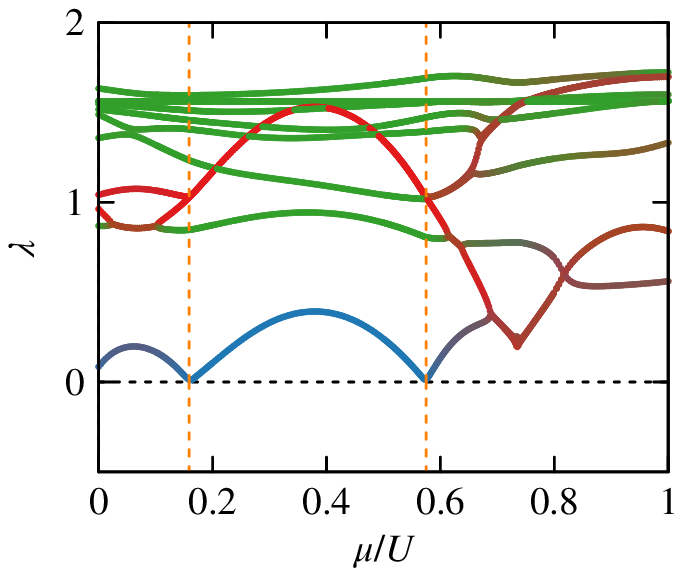}
	\caption{
		(left) Spectra of the matrices $g_{2}^{(0)}$ and $g_{01}^{(0)}$ for all valid $y$ as a function of $\mu/U$,
		obtained for $M = 5$, $T/U = 0.08$, $J_{0}/U = 0.122$, and $J/U = 0.1$.
		The vertical dashed lines mark the phase transition points
		between the disordered (between the two lines) and superfluid (outside the lines) phases.
		The red arrow marks the point where a negative eigenvalue appears.
		(right) Spectrum of the matrix $g_{01}^{(0)}$ repeated from panel (a).
		Colors of the lines denote the composition of the eigenvectors associated with the eigenvalues:
		red, green, and blue color components show the size of the eigenvector projection onto subspaces
		spanned by $\lbrace\eta, \xi\rbrace$, $\lbrace\epsilon, \zeta\rbrace$, and $\lbrace\rho\rbrace$, respectively.
	}
	\label{fig:lines-J10}
\end{figure}
Based on the cut (A) analysis,
we focus only on the eigenvalues of $g_{2}^{(0)}$ and $g_{01}^{(0)}$ for the remaining cuts.
First, we look at the cut (B) depicted in Fig.~\ref{fig:lines-m4}, where there is a disordered phase up to $J/U\approx 0.127$,
while above $J/U\approx 0.127$, there is a glass phase.
Again $g_{2}^{(0)}$ has negative eigenvalues,
but this time the point where the instability occurs coincides with the disordered-glass transition point.
At the same point, one eigenvalue of $g_{01}^{(0)}$ reaches zero, but it does not go below.
The third cut, (C), shown in Fig.~\ref{fig:lines-J10},
contains the disordered phase between $0.16 \lesssim \mu/U \lesssim 0.57$
and the superfluid phase below and above these values.
However, again, the matrix $g_{01}^{(0)}$ has a zero eigenvalue at the phase transition,
while the negative eigenvalue comes from the matrix $g_{2}^{(0)}$,
splitting the superfluid phase into a stable region below $\mu/U\approx 0.75$ and an unstable one above it.

\section{Final stability criterion}\label{sec:criterion}
Based on the analysis of the spectra of all considered matrices across various parameter values,
we conclude that the instability can be found by checking whether the matrix $g_{2}^{(0)}$ has negative eigenvalues.
Thus, to check the stability of the solution in the current model,
one needs to evaluate a matrix of the form (repeated from Eq.~\eqref{eq:effmat-2} for convenience)
\begin{equation}\label{eq:g2-0}
	g_{2}^{(0)} = \mbox{$\footnotesize
	\begin{pmatrix}
		\mathbb{X''}-2\mathbb{X'}+\mathbb{X} & \mathbb{Y''}-2\mathbb{Y'}+\mathbb{Y}\\
		\mathbb{Y''}-2\mathbb{Y'}+\mathbb{Y} & \mathbb{Z''}-2\mathbb{Z'}+\mathbb{Z}\\
	\end{pmatrix}$},
\end{equation}
where
{\small
\begin{align}
\nonumber&\mathbb{X''} = \sum_{l,d}\frac{\partial^2 \mathcal{F}}{\partial \eta_{\alpha\beta,k,k'} \partial \eta_{\alpha\beta, k+l,k'+d}}, &
		 &\mathbb{X'}  = \sum_{l,d}\frac{\partial^2 \mathcal{F}}{\partial \eta_{\alpha\beta,k,k'} \partial \eta_{\alpha\gamma,k+l,k'+d}}, &
		 &\mathbb{X}   = \sum_{l,d}\frac{\partial^2 \mathcal{F}}{\partial \eta_{\alpha\beta,k,k'} \partial \eta_{\gamma\delta,k+l,k'+d}}, \\
\nonumber&\mathbb{Y''} = \sum_{l,d}\frac{\partial^2 \mathcal{F}}{\partial \eta_{\alpha\beta,k,k'} \partial  \xi_{\alpha\beta, k+l,k'+d}}, & 
		 &\mathbb{Y'}  = \sum_{l,d}\frac{\partial^2 \mathcal{F}}{\partial \eta_{\alpha\beta,k,k'} \partial  \xi_{\alpha\gamma,k+l,k'+d}}, & 
		 &\mathbb{Y}   = \sum_{l,d}\frac{\partial^2 \mathcal{F}}{\partial \eta_{\alpha\beta,k,k'} \partial  \xi_{\gamma\delta,k+l,k'+d}}, \\
		 &\mathbb{Z''} = \sum_{l,d}\frac{\partial^2 \mathcal{F}}{\partial  \xi_{\alpha\beta,k,k'} \partial  \xi_{\alpha\beta, k+l,k'+d}}, & 
		 &\mathbb{Z'}  = \sum_{l,d}\frac{\partial^2 \mathcal{F}}{\partial  \xi_{\alpha\beta,k,k'} \partial  \xi_{\alpha\gamma,k+l,k'+d}}, & 
		 &\mathbb{Z}   = \sum_{l,d}\frac{\partial^2 \mathcal{F}}{\partial  \xi_{\alpha\beta,k,k'} \partial  \xi_{\gamma\delta,k+l,k'+d}}, 
\end{align}
}%
where $\alpha\neq\beta\neq\gamma\neq\delta$, $k$ and $k'$ may be chosen arbitrarily,
while the derivatives can be worked out using Eq.~\eqref{eq:deri}.
If this matrix has negative eigenvalues, the solution is not stable.

\section{Summary}\label{sec:summary}
We have studied the stability of the replica-symmetric solution for a system of strongly interacting bosons with off-diagonal disorder.
We have derived a set of conditions necessary for the solution to be stable
and identified one of them as the strongest one in all of the considered parameter choices.
It is an analogue of the ``$P-2Q+R$'' eigenvalue~\cite{Almeida1978_JoPAMaG11},
which was found to be the lowest near the transition point for the spin-glass systems.
The simplified form of the condition derived by us can be found in Eq.~\eqref{eq:g2-0} and its description.

We have also evaluated this condition for selected values of the system parameters,
to determine the stability of the various phases present in the system.
In addition to the expected stability of the disordered phase and instability of the glass phase,
we have found that the superfluid phase consists of a stable and unstable part.
We may recognize these two parts as a clean superfluid and a superglass phase~\cite{Piekarska2022}, respectively,
which we, however, do not elaborate on here, as it is not the goal of the current paper.

A full description of the unstable phases and the transitions between them requires breaking the stability of the replicas.
This is, however, a very demanding task, as even the replica-symmetric derivation is already rather complex
and on the verge of feasibility in terms of the numerical computation times.

\bibliography{all}
\bibliographystyle{apsrev4-2}
\end{document}